\documentclass[10pt]{article}
\usepackage[utf8]{inputenc}
\usepackage[T1]{fontenc}
\usepackage{graphicx}
\usepackage{grffile}
\usepackage{longtable}
\usepackage{wrapfig}
\usepackage{rotating}
\usepackage[normalem]{ulem}
\usepackage{amsmath}
\usepackage{textcomp}
\usepackage{amssymb}
\usepackage{capt-of}
\usepackage{hyperref}
\usepackage{listings}
\usepackage{vub}
\usepackage{listings}
\usepackage{color}
\usepackage{placeins}
\usepackage{biblatex}
\faculty{Sciences and Bio-Engineering Sciences}
\author{Sander Huyghebaert \linebreak Thomas Van Strydonck \linebreak Dr. Steven Keuchel \linebreak Prof. Dr. Dominique Devriese}
\date{2019-2020}
\title{Uninitialized Capabilities\\\medskip
\large Technical Report}
\hypersetup{
 pdfauthor={Sander Huyghebaert \linebreak Thomas Van Strydonck \linebreak Dr. Steven Keuchel \linebreak Prof. Dr. Dominique Devriese},
 pdftitle={Uninitialized Capabilities},
 pdfkeywords={},
 pdfsubject={},
 pdfcreator={Emacs 26.3 (Org mode 9.1.9)}, 
 pdflang={English}}
\begin{document}

\maketitle
\setcounter{tocdepth}{2}
\tableofcontents

\newpage

\section{Introduction}
\label{sec:orgb8044f5}
This technical report describes a new extension to capability machines. Capability machines
are a special type of processors that include better security primitives at the hardware level.
In capability machines, every word has an associated tag bit that indicates whether
the value it contains is a capability or a regular data value. Capabilities enable fine-grained
control of the authority over memory that program components have. Conceptually, capabilities
can be viewed as being an unforgeable token carrying authority over a resource.

CHERI \parencite{watson2019capability} is a recently developed capability machine that aims to provide
fine-grained memory protection, software compartmentalization and backwards compatibility. While 
our ideas are implemented on CHERI, they are not limited to it and should be applicable to other
capability machines as well.

In this technical report we propose a new type of capabilities, which represent the authority to 
access (read and write to) a block of memory but not view its initial contents. Our main goal is to 
use this new type of capability as part of a secure calling convention, but other applications may be possible too.

\section{Capability Machines}
\label{sec:org4f22d22}
Capability machines are a special type of processor that replaces pointers with capabilities.
Conceptually, capabilities are tokens that carry authority to access memory or an object. When
capabilities represent software defined authority like invoking objects or closures, they are referred to as \emph{object capabilities}. 
This technical report will focus on primitive capabilities for accessing memory.
The permissions to access memory can be read only, read and write to, execute, \ldots{} 
The idea of capabilities was first formally defined by Dennis and Van Horn \parencite{dennis1966programming} 
and has been further explored in the decades after. 

The first capability machine dates back from 1959 with the \emph{Rice University Computer} and the 
development and research interest of capability machines slowed down significantly after the 
\emph{iAPX 432} from \emph{Intel} in 1981 \parencite{levy2014capability}. 
In 2014, researchers of the University of Cambridge developed a new 
capability machine: \emph{CHERI}, on which we will provide an implementation of our contribution to
capability machines.

It is important that capabilities cannot be forged, as forging them with certain permissions, 
memory bounds, etc. would defeat their purpose. One of the solutions to
ensure the unforgeability of capabilities is to provide specialized instructions to work
with capabilities. Capabilities might however need to be stored in primary memory or
secondary memory instead of just the registers on the processor and one of the most used 
techniques to ensure valid capabilities is the use of tagged memory \parencite{fabry1974capability}. 
Every possible capability location will have a tag denoting if that location contains a capability 
or not. Capabilities for which the tag is not set cannot be used to dereference memory.

Some common permissions found on capability machines are:
\begin{itemize}
\item \textbf{R}: read-only;
\item \textbf{RW}: read-write;
\item \textbf{RX}: read-execute;
\item \textbf{RWX}: read-write-execute.
\end{itemize}

For this technical report we will represent capabilities formally as a 4-tuple similar to the
representation used by Skorstengaard et al. \parencite{skorstengaard2018reasoning},
(\emph{permissions}, \emph{base}, \emph{end}, \emph{cursor}), this tuple contains the 
\emph{permissions} of the capability, the range to which these permissions apply \([base, end]\) and 
a \emph{cursor} in that range. 

\section{Uninitialized Capabilities}
\label{sec:org05a85db}
\begin{figure}[htbp]
\centering
\includegraphics[width=0.5\textwidth]{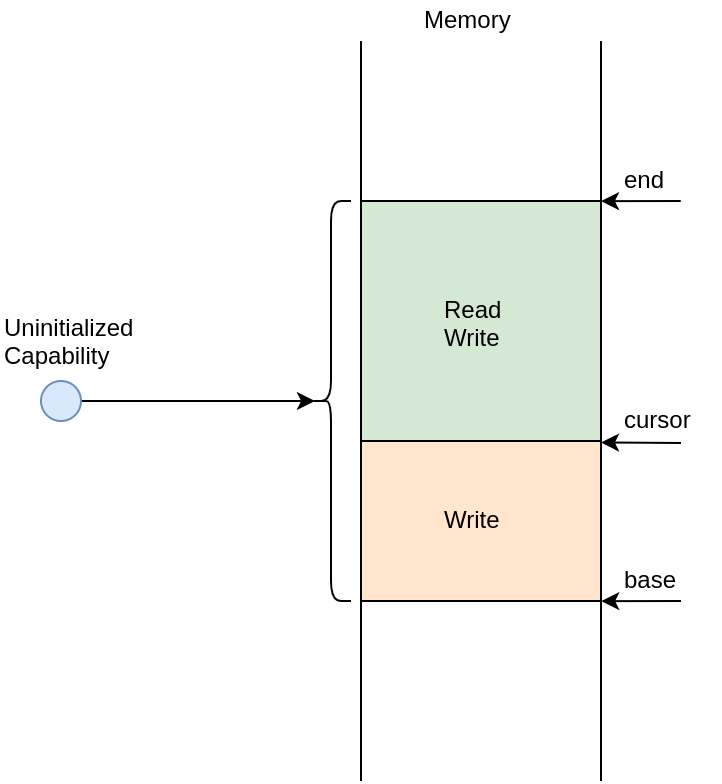}
\caption{\label{fig:uninit-cap-concept}Uninitialized Capabilities Concept}
\end{figure}

Uninitialized capabilities are a new type of capabilities.
They are memory capabilities which represent read-write authority to a range of memory, except that they do not allow reading the initial contents of the memory.
The memory first needs to be overwritten before it can be read.
This type of capability requires a new permission to be added to capabilities (\textbf{U}: uninitialized) and prevents the holder of the capability from reading memory that they have not first initialized.
Figure \ref{fig:uninit-cap-concept} clarifies this concept a bit more.

\FloatBarrier

Formally, uninitialized capabilities grant the following authority:
\begin{itemize}
\item permission to read in \([cursor, end]\);
\item permission to write in \([base, end]\);
\item when writing immediately below the cursor, the cursor will be decremented so that the holder of the 
uninitialized capability is able to read from the location it has just written to.
\end{itemize}

Uninitialized capabilities can thus be used to give access to arrays that contain uninitialized 
data without the need for clearing that uninitialized data first.

The full set of permissions becomes:
\begin{itemize}
\item \textbf{R}: read-only;
\item \textbf{RW}: read-write;
\item \textbf{RX}: read-execute;
\item \textbf{RWX}: read-write-execute.
\item \textbf{U}: read between \([cursor, end]\), write between \([base, end]\);
\end{itemize}

We have chosen not to include combinations of the \textbf{U} permission and \textbf{X} permission. 
Executing an uninitialized capabilities would require
incrementing the program counter (and thus the cursor of the uninitialized capability),
which means that the non-readable range of the capability would grow.

Another option is to allow the combination of the \textbf{U} permission with the \textbf{X} permission, 
but when jumping to an uninitialized capability transform it into a normal capability for
the range \([cursor, end]\) before placing it in the program counter capability register.

We propose a concrete design of uninitialized capabilities for the \emph{CHERI} capability machine, particularly the CHERI-MIPS ISA.
However, the general concept is not limited to CHERI-MIPS.
We see the concept of uninitialized capabilities as an addition to capability machines in general, and particularly the CHERI protection model, regardless of the architecture it is run on.

\subsection{CHERI}
\label{sec:orgd6045be}
CHERI (\textbf{C}apability \textbf{H}ardware \textbf{E}nhanced \textbf{R}ISC \textbf{I}nstructions) 
is an ISA extension that introduces capabilities. The main goals of CHERI are 
fine-grained memory protection, software compartmentalization and backwards compatibility \parencite{watson2019capability}.

The CHERI ISA extension proposes a 64-bit, 128-bit and 256-bit capability representation format \parencite{watson2019capability}, 
we instantiate our ideas for the 256-bit capability format but it should be possible to 
add the uninitialized permission bit to other formats as well.

In Figure \ref{fig:cap-256} we see the current 256-bit capability format:

\begin{figure}[htbp]
\centering
\includegraphics[width=0.8\textwidth]{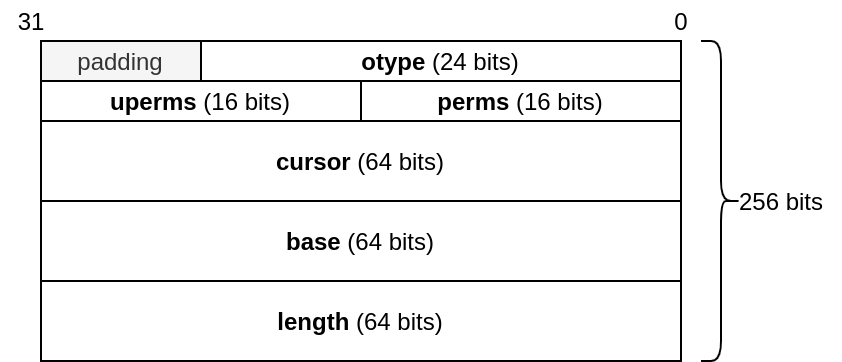}
\caption{\label{fig:cap-256}256-bit Capability Representation Format}
\end{figure}
\FloatBarrier

The important fields of a capability for our proposal are the permissions, cursor, base and
length fields. In our formal representation of capabilities we have an \emph{end} field instead of
\emph{length} but it should be straightforward to see that \(end = base + length\).

In the next section, we instantiate uninitialized capabilities as a set of modifications/additions to the CHERI-MIPS ISA.
We have implemented these for CHERI-MIPS in software (using a simulator).

\subsection{Uninitialized Capabilities Implementation}
\label{sec:org0980643}
\subsubsection{Uninitialized Permission Bit}
\label{sec:orgd0efaa5}
\begin{figure}[htbp]
\centering
\includegraphics[width=0.8\textwidth]{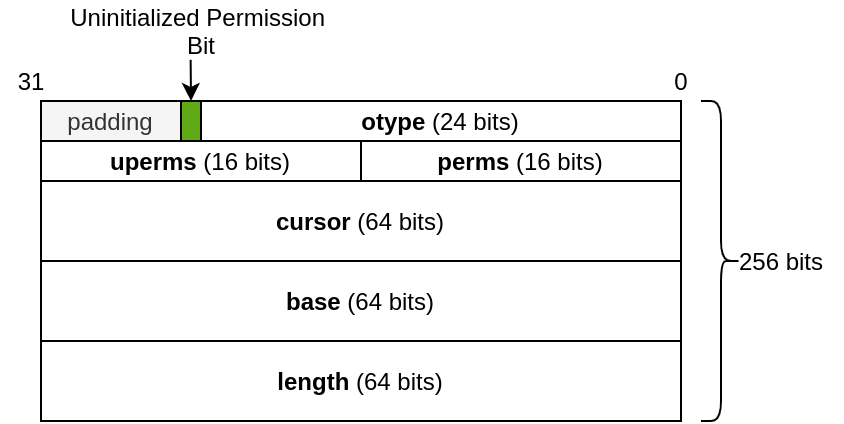}
\caption{\label{fig:uninit-cap-rep}Modified 256-bit representation of a capability}
\end{figure}

The first modification that needs to be made to CHERI capabilities is the addition of a new
permission, the uninitialized permission. In the 256-bit capability format there are a few 
unused bits (padding bits) available so we have opted to use one of those bits for the 
uninitialized permission, as can be seen in Figure \ref{fig:uninit-cap-rep}.

\FloatBarrier

\subsubsection{Instruction Modifications}
\label{sec:orgc892f59}
A few instruction were modified to take the uninitialized permission into account. What follows
is a list of the instructions modified and a description of what that modification entails:

\bigskip
\noindent
\textbf{Load via Capability Register (CL[BHWD][U]/CLC)}: When load instructions are given a capability
with the uninitialized permission set, it is not allowed to load from an address lower
than the cursor.

\bigskip
\noindent
\textbf{Set/Increment Offset Or Address (CSetOffset/CIncOffset/CIncOffsetImm/CSetAddr/CAndAddr)}: Instructions that modify 
the cursor of an uninitialized capability are not allowed to set the cursor lower than it originally 
was. The only way to lower the cursor is by using the uninitialized store instructions.

\subsubsection{New Instructions}
\label{sec:orgecbb5b4}
We propose new instructions for the implementation of uninitialized capabilities:

\bigskip
\noindent
\textbf{Get Uninitialized Bit of a Capability (CGetUninit)}: This instructions has 2 parameters,
the general-purpose register to store the uninitialized bit of the capability into and
the capability of which the uninitialized bit is requested.

\bigskip
\noindent
\textbf{Uninitialize a Capability (CUninit)}: An instruction to make a capability uninitialized.
This instruction takes a source capability register and a destination capability register that
will contain the capability from the source register but with the uninitialized permission set.
An error will be generated if the original capability did not have read-write authority.

\bigskip
\noindent
\textbf{Uninitialized Store (UCS[BHWD]/UCSC)}: These instructions are modified versions of their
not-uninitialized counterparts (CS[BHWD], CSC).
They behave similarly to the normal store instructions, except when the given offset is \(-1\) and the capability used for the store is uninitialized.
In that case, the capability
written to the destination capability register will have the cursor of the source capability 
decremented by the number of bytes written (i.e. 1 for a byte, 2 for a half word, 4 for a word,
8 for a double word and 32 for capabilities when using the 256-bit capability format). 
Specifying an offset of \(-1\) is the \textbf{only} way to decrement the cursor.
This instruction takes 4 arguments, a destination capability register (which will contain
the source capability but possibly with its cursor modified if the offset was \(-1\)), a source 
register for the data to write, an offset and a source capability register.

\bigskip
The original store instructions for capabilities are \textbf{not} modified (CSC, CSW, \ldots{}), but instead 
we propose to add new instructions to handle the uninitialized permission. The new instructions
write to a capability register the possibly modified capability (if it has the \textbf{U} permission
set and the given offset is \(-1\)), while the original instructions do not write to a register but
instead allow specifying a register containing another offset to be added to the cursor of the
capability.

One additional instruction is required to modify the bounds of uninitialized capabilities:

\bigskip
\noindent
\textbf{Shrink a Capability (CShrink[Imm])}: CShrink is an instruction with 3 parameters, the destination
capability register, the source capability register and a general-purpose register (GPR), or alternatively
an unsigned immediate for CShrinkImm. The capability from the source register will be 
modified by setting \(end = cursor\) and \(base = value\ in\ GPR\) for CShrink. For CShrinkImm
\(end = cursor\) and \(base = base + immediate\). CShrink[Imm] will raise an exception if the
\(end < cursor\) (the original \(end\) and \(cursor\) of the capability) or if \(newBase < base\), 
these conditions prevent expanding the range of authority of the capability.

\bigskip
In the CHERI-MIPS ISA a similar instruction is already available, \textbf{CSetBounds}, but this 
instruction did not meet the needs of uninitialized capabilities. It adjusts the bounds
by setting \(base = cursor\) and \(end = cursor + immediate\), where immediate is either the value
from the general-purpose register specified in the instruction or an unsigned immediate value.

The issue with using this instruction in combination with uninitialized capabilities arises
when trying to lower the \(end\) of the uninitialized capability, but maintain the same \(base\).
Using CSetBounds this would require first setting \(cursor = base\), calculate the offset
for the new \(end\), perform the CSetBounds instruction and then setting the \(cursor\) back
to its value before it was set to \(base\). This obviously means lowering the cursor (\(cursor = base\))
which is not permitted for uninitialized capabilities.

\section{Secure Calling Convention}
\label{sec:org5bb3ea3}
In the paper "Reasoning About a Machine with Local Capabilities" \parencite{skorstengaard2018reasoning},
a novel calling convention is proposed by using local capabilities. This calling convention ensures
local stack frame encapsulation and well bracketed control flow on a single shared stack. 
We propose to continue using this calling convention with the slight modification that the stack
capability should be made uninitialized on function invocation.

The calling convention \parencite{skorstengaard2018reasoning} mentions as a point of improvement
to the calling convention the need for an instruction for efficiently clearing a large part of 
memory. Uninitialized capabilities can be used to prevent this overhead of clearing 
(by making the stack capability uninitialized). 
Note that it is still necessary that a callee clears its used stack frame.

\begin{figure}[htbp]
\centering
\includegraphics[width=0.8\textwidth]{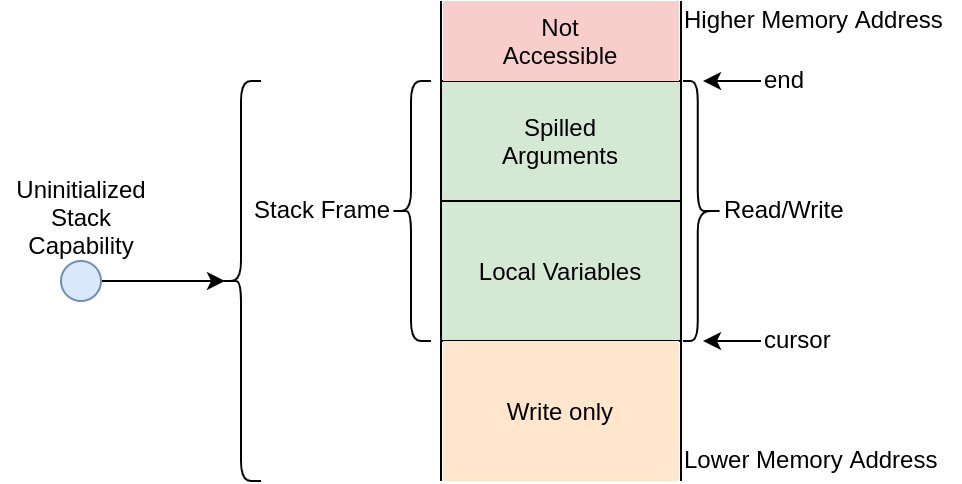}
\caption{\label{fig:uninit-stack}Stack with Uninitialized Capability}
\end{figure}

Having an uninitialized stack capability prevents adversaries from reading from the stack unless they first 
overwrite the uninitialized data (this could be garbage but also sensitive data or capabilities that they should not get access to).
See Figure \ref{fig:uninit-stack} for a conceptual diagram of having an uninitialized stack
capability.

\FloatBarrier

\section{Conclusion}
\label{sec:orgd41147b}
We have proposed a new permission of capabilities, the \textbf{uninitialized} permission. This permission
only allows reading those parts of memory denoted by the bounds of the capability to which it
has first written to. This prevents using the capability to read uninitialized data (be it garbage
or sensitive data). We also provided a brief discussion of how to implement this on the \emph{CHERI}
capability machine. Finally we showed how uninitialized capabilities can contribute to
secure calling conventions.

\newpage
\printbibliography
\end{document}